\documentclass[twocolumn,twoside,slac_two]{revtex4}
\usepackage{graphicx}
\usepackage{fancyhdr}
\usepackage{natbib}
\def\be{\begin{equation}}
\def\ee{\end{equation}}

\begin{document}

\title{Gamma-Ray Pulsar Light Curves in Vacuum and Force-Free Geometry}

%

\author{Alice K. Harding$^1$, Megan E.  DeCesar$^{1,2}$, M. Coleman Miller$^{2}$, Constantinos Kalapotharakos$^{1,2}$, Ioannis Contopoulos$^{3}$}
\affiliation{$^1$Astrophysics Science Division,  NASA Goddard Space Flight Center, Greenbelt, MD 20771, USA}
\affiliation{$^2$Department of Astronomy, University of Maryland, College Park, MD 20742, USA}
\affiliation{$^3$Research Center for Astronomy, Academy of Athens, Athens 11527, Greece}

\begin{abstract}
Recent studies have shown that gamma-ray pulsar light curves are very sensitive to the geometry of the pulsar magnetic field.  Pulsar magnetic field geometries, such as the retarded vacuum dipole and force-free magnetospheres have distorted polar caps that are offset from the magnetic axis in the direction opposite to rotation.  Since this effect is due to the sweepback of field lines near the light cylinder, offset polar caps are a generic property of pulsar magnetospheres and their effects should be included in gamma-ray pulsar light curve modeling.  In slot gap models (having two-pole caustic geometry), the offset polar caps cause a strong azimuthal asymmetry of the particle acceleration around the magnetic axis.  We have studied the effect of the offset polar caps in both retarded vacuum dipole and force-free geometry on the model high-energy pulse profiles.  We find that, compared to the profiles derived
from symmetric caps, the flux in the pulse peaks, which are caustics formed along the trailing magnetic field lines, increases significantly relative to the off-peak emission, formed along leading field lines.   The enhanced contrast produces improved slot gap model fits to {\it Fermi} pulsar light curves like Vela, 
with vacuum dipole fits being more favorable.
\end{abstract}

\maketitle

\thispagestyle{fancy}


\section{INTRODUCTION}

The Fermi Gamma-Ray Space Telescope has had a major impact on pulsar physics with the discovery of over 100 gamma-ray pulsars comprising three  
populations: young radio-loud pulsars, young radio-quiet pulsars and millisecond pulsars \cite{Smith2011}.  The wide variety of light curve types, viewing 
geometry and spin-down luminosity encompassed by this large group of sources provides an unprecedented opportunity to explore and constrain 
pulsar emission and magnetic field geometry.  There have also been significant advances recently in numerical simulation of pulsar magnetosphere models 
\cite{Spitkovsky2006}\cite{ContKala2010}\cite{Kala2011}\cite{Li2011}  that define the high-altitude magnetic field structure critical to the emission gap models.  Since the radiation  
in the outer magnetosphere high-energy emission models, such as outer gap \cite{RomaniYadigaroglu1995} and slot gap \cite{MuslimovHarding2004}, occurs along the 
boundary of the open field region, the predicted gamma-ray light curves are sensitive to the magnetic field structure \cite{RomaniWatters2010}.  
In particular, the sweepback of the magnetic 
field lines near the light cylinder, due to retardation and currents, causes an offset of the polar cap in the direction opposite to the rotation.  In the slot gap (SG) model,
the polar cap offset produces two main effects on the emission and the resulting light curves: an azimuthal asymmetry in the SG emission, and a change in the phase 
lag between the first gamma-ray peak and the radio peak.
We present a study of SG model light curves for the two extreme cases of vacuum dipole and force-free pulsar magnetospheres to examine such effects.

\section{PULSAR MAGNETOSPHERE MODELS}

The global structure of a pulsar magnetosphere is a presently unsolved problem.  An analytic expression exists only for the structure
of the magnetosphere of the vacuum rotator (the Deutsch solution, \cite{Deutsch1955}), but real pulsars are certainly not in vacuum since electrons and
positrons are copiously produced due to the high surface electric fields induced by rotation.  The other extreme, known as force-free 
models \cite{Contopoulos1999}\cite{Spitkovsky2006}\cite{Timokhin2006} that specify the configuration of fields, charge and currents that 
solve Maxwell's Equations in the approximation of ideal-MHD (${\bf E \cdot B} = 0$) neglecting plasma inertia,  is also well determined.
Models with finite conductivity that bridge these two extremes have been studied very recently \cite{Kala2011}\cite{Li2011}.     
These simulations show that there is a range of magnetic field structures, 
current distributions and spin-down power that lie between the vacuum dipole, with 
maximum accelerating electric field but no charges, and the force-free solutions, with charges and current but no acceleration. 

\begin{figure*}[t]
\centering
\includegraphics[width=135mm]{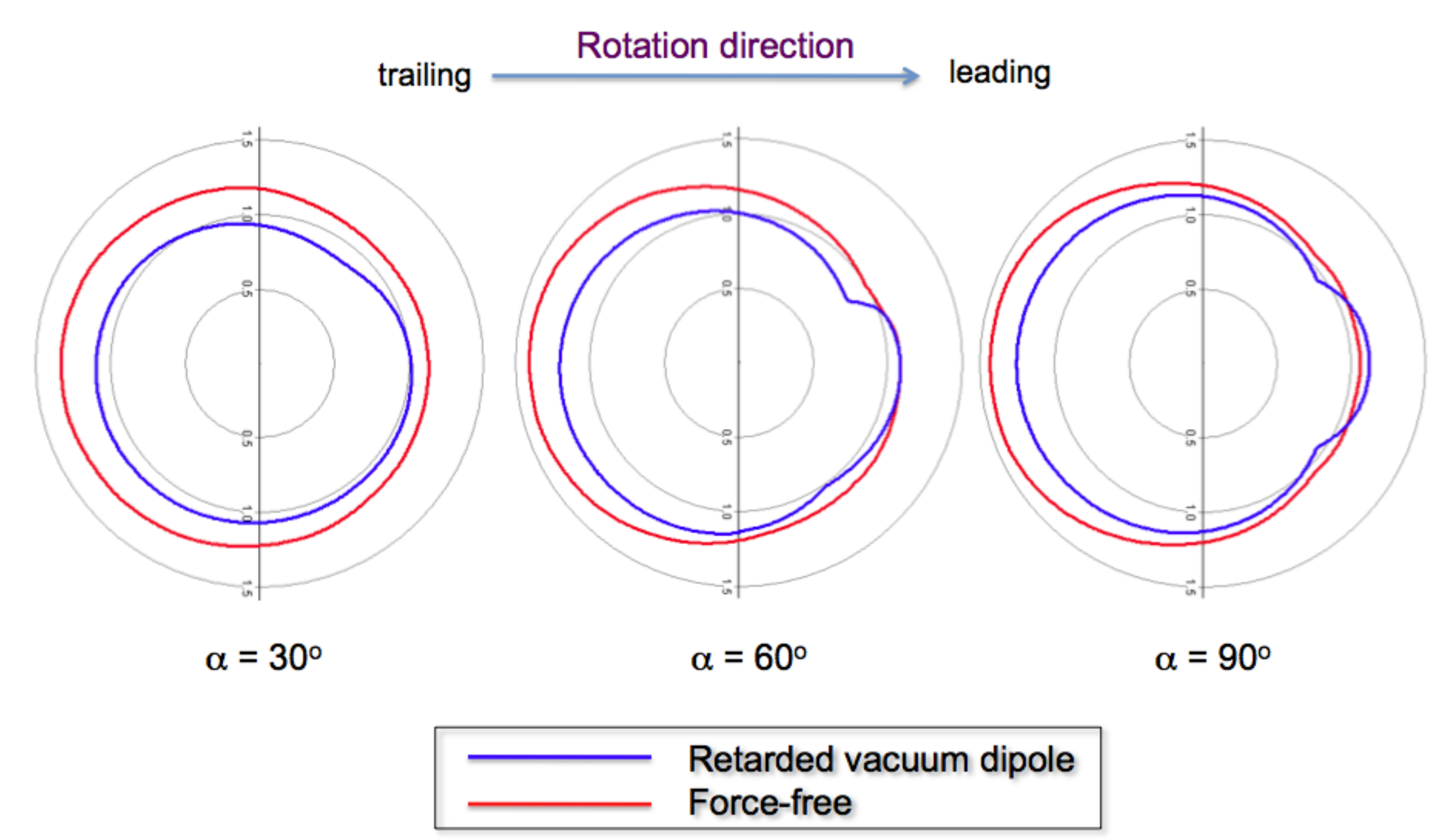}
\caption{Shape of polar caps for retarded vacuum dipole (blue) and force-free (red) magnetospheres for different pulsar inclination angles $\alpha$. } \label{f1}
\end{figure*}

One property that all these solutions have in common is sweepback of the magnetic field lines near the light cylinder opposite to the direction of rotation, as 
the poloidal dipole field transitions into the pulsar wind (or electromagnetic wave in the vacuum case).  The magnetospheres of all these solutions divide into open 
field lines, those which cross the light cylinder, and closed field lines, those that close within the light cylinder.  
Tracing the bundle of open field lines to their footpoints on the neutron star surface defines the polar cap (PC).  Due to the sweepback, the PCs are asymmetric with 
their centers shifted backward in phase relative to the dipole axis \cite{ArendtEilek1998}\cite{DyksHarding2004}.  Examples of such offset PCs are shown in Figure 1 for the
retarded vacuum and force-free models.  One can see that the PC rims in both models are shifted toward the trailing side, the amount of shift being a function of inclination 
angle, with the force-free model PCs shifted significantly 
more than the those of the vacuum model (see also \cite{BaiSpitkovsky2010}).  In addition, the force-free PCs are larger than those of the vacuum, especially for small inclination,
because the volume of open field is larger.  The notches that appear in vacuum PC rims for the higher inclination angles largely disappear in the force-free model PCs.

\section{MODEL LIGHT CURVES FOR OFFSET POLAR CAPS}

To explore how the magnetic field structure and offset PCs influence gamma-ray pulsar light curves, we have generated model light curves using a geometrical 
version of the SG, often referred to as the two-pole caustic (TPC) model \cite{DyksRudak2003}.  The physical SG has its origin in polar cap pair cascades that screen the 
accelerating parallel electric field $E_{\parallel}$ above a pair formation front (PFF) \cite{Arons1983}\cite{HardingMuslimov1998}.  Approaching the last open field line, assumed 
to be a perfectly conducting boundary where $E_{\parallel}$ vanishes, the PFF moves to higher altitude as particles must accelerate over a longer distance to radiate 
pair-producing photons.  A narrow gap of unscreened $E_{\parallel}$ forms along this boundary, encompassing field lines along which the potential drop necessary
 for pair production is never achieved.  The electrons in the SG keep accelerating to high altitude, radiating curvature and possibly synchrotron and inverse Compton 
 emission\cite{MuslimovHarding2004}\cite{Harding2008}.  

\subsection{Asymmetric Slot Gap Emission}

\begin{figure*}[t]
\centering
\includegraphics[width=150mm]{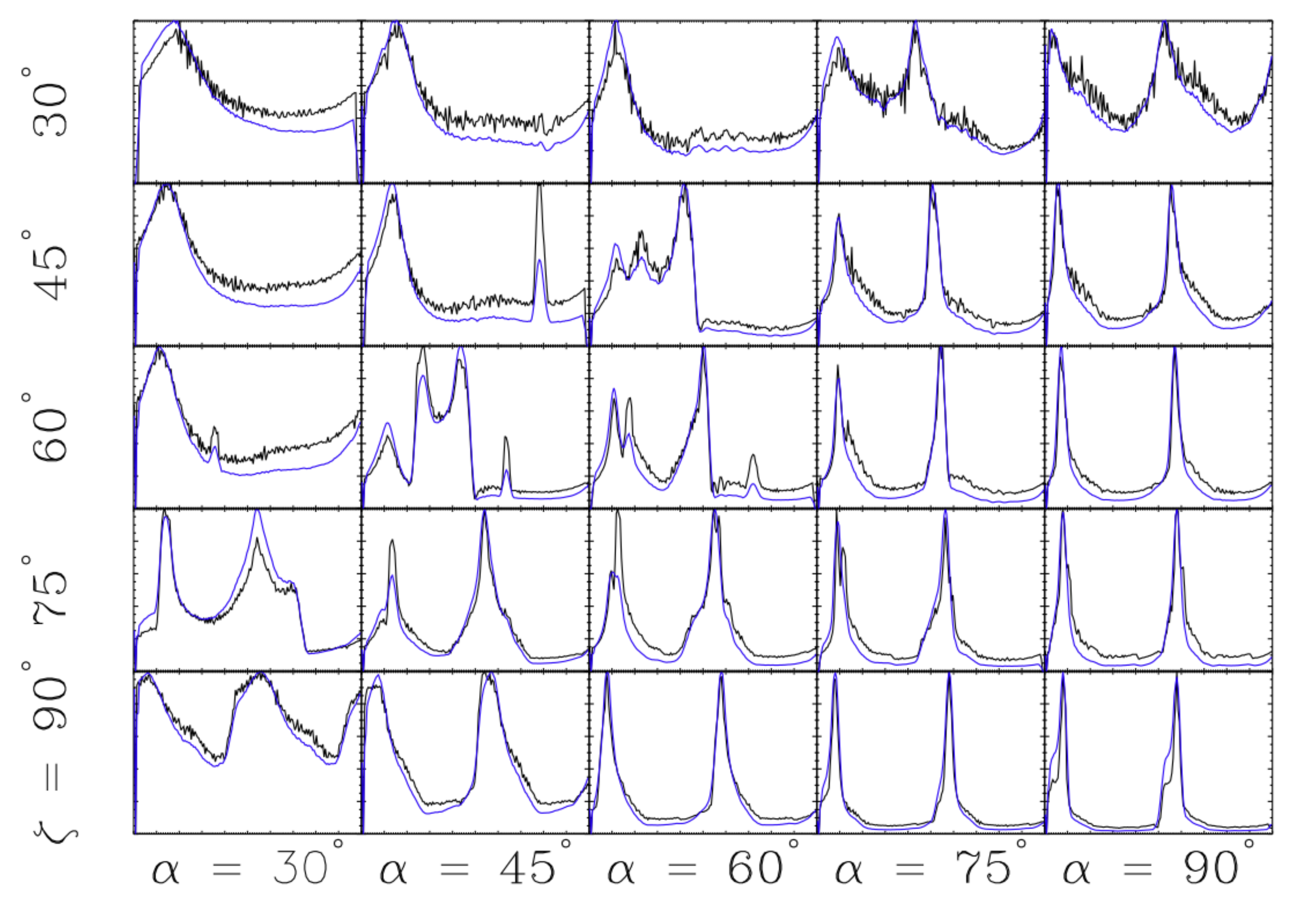}
\caption{Simulated gamma-ray light curves for different combinations of pulsar inclination angle $\alpha$ and viewing angle $\zeta$ with respect to the rotation axis, 
in a retarded vacuum dipole magnetosphere for symmetric (black) and asymmetric (blue) slot gap geometry.  The assumed slot gap width is 0.05 of the open field 
lines and the maximum emission radius is $1.2 R_{\rm LC}$.  The phase of closest approach to the magnetic pole is at 0.} \label{f2}
\end{figure*}

The original SG model $E_{\parallel}$ \cite{Arons1983} had azimuthal asymmetry because field lines curved toward the rotation axis ${\bf \Omega}$ (`favorably curved') 
accelerate one sign of charge (electrons in the ${\bf \Omega \cdot B} > 0$ case) and field lines curved away from the rotation axis (`unfavorably curved') accelerate the other 
sign of charge.  Inclusion of frame-dragging \cite{MuslimovTsygan1992} introduced a larger favorable $E_{\parallel}$ contribution that significantly reduced this azimuthal 
asymmetry.  Both of these $E_{\parallel}$ solutions assumed centered dipole PCs.  Recently Harding \& Muslimov \cite{HardingMuslimov2011a}\cite{HardingMuslimov2011b} 
presented a solution for 
$E_{\parallel}$ in non-dipolar magnetic fields having PCs with offsets of arbitrary degree and direction.  
\be
{\bf B} \approx {B_0\over {\eta ^3}}~\left[ \hat{\bf r}~\cos [\theta (1+a)] +{1\over 2} ~\hat{\bf \theta}~ \sin [\theta (1+a)]  \right],
\label{B}
\ee
where $B_0$ is the surface magnetic field strength at the magnetic pole,  $\eta = r/R$ is the dimensionless radial coordinate in units of stellar radius, $R$,
$a=\varepsilon ~\cos (\phi - \phi _0)$ is the parameter characterizing the distortion of polar field lines, and $\phi_0$ is the magnetic azimuthal angle defining 
the meridional plane of the offset PC.  In this magnetic field, the PC angle has an azimuthal dependence, 
\be \label{theta0}
\theta_0 \approx {1 \over {1+a}}~\sin ^{-1}\left(x^{{1\over 2}(1+a)} \right), 
\ee
respectively, where $x = r/R_{\rm LC}$ is the radial distance in units of the light cylinder radius, $R_{\rm LC} = c/\Omega$.
Since the $E_{\parallel}$ has a dependence on the PC angle, its value is significantly larger on the offset side of the PC, producing a strong azimuthal asymmetry.   

If we assume that ${\bf \Omega}$ is along the z-axis and that the magnetic axis lies in the x-z plane, then the PC offsets due to magnetic field sweepback are in the y-z plane 
($\phi _0 = \pi /2$), so that $a = \varepsilon \sin \phi $.  In this case, the $E_{\parallel}$ of the SG will have an approximate dependence \cite{HardingMuslimov2011b}
\be  \label{Epar}
E_{\parallel} (a) \approx {\theta_0^{2a}\over(1+a)^2}\,E_{\parallel}(0)
\ee
where $E_{\parallel}(0)$ is the value for the case of a pure dipole field.  We can quantify the offset parameter $\varepsilon$ in different magnetosphere models by the 
relation
\be
\theta_0^{{-\varepsilon}} \approx {R_{\rm tr}\over R_{\rm PC}},
\ee
where $R_{\rm tr}$ is the radius of the trailing side of the PC and $R_{\rm PC} \sim R(\Omega R /c)^{1/2}$ is the radius of a centered PC.  We find that for the retarded vacuum
dipole, $\varepsilon \sim 0.05 - 0.1$ and for the force-free model, $\varepsilon \sim 0.1 - 0.2$, with minimum and maximum corresponding to inclination angles of $0^\circ$ 
and $90^\circ$ respectively.  
 
\subsection{Light Curves in Vacuum and Ideal Force-Free Magnetospheres}

\begin{figure*}[t]
\centering
\includegraphics[width=150mm]{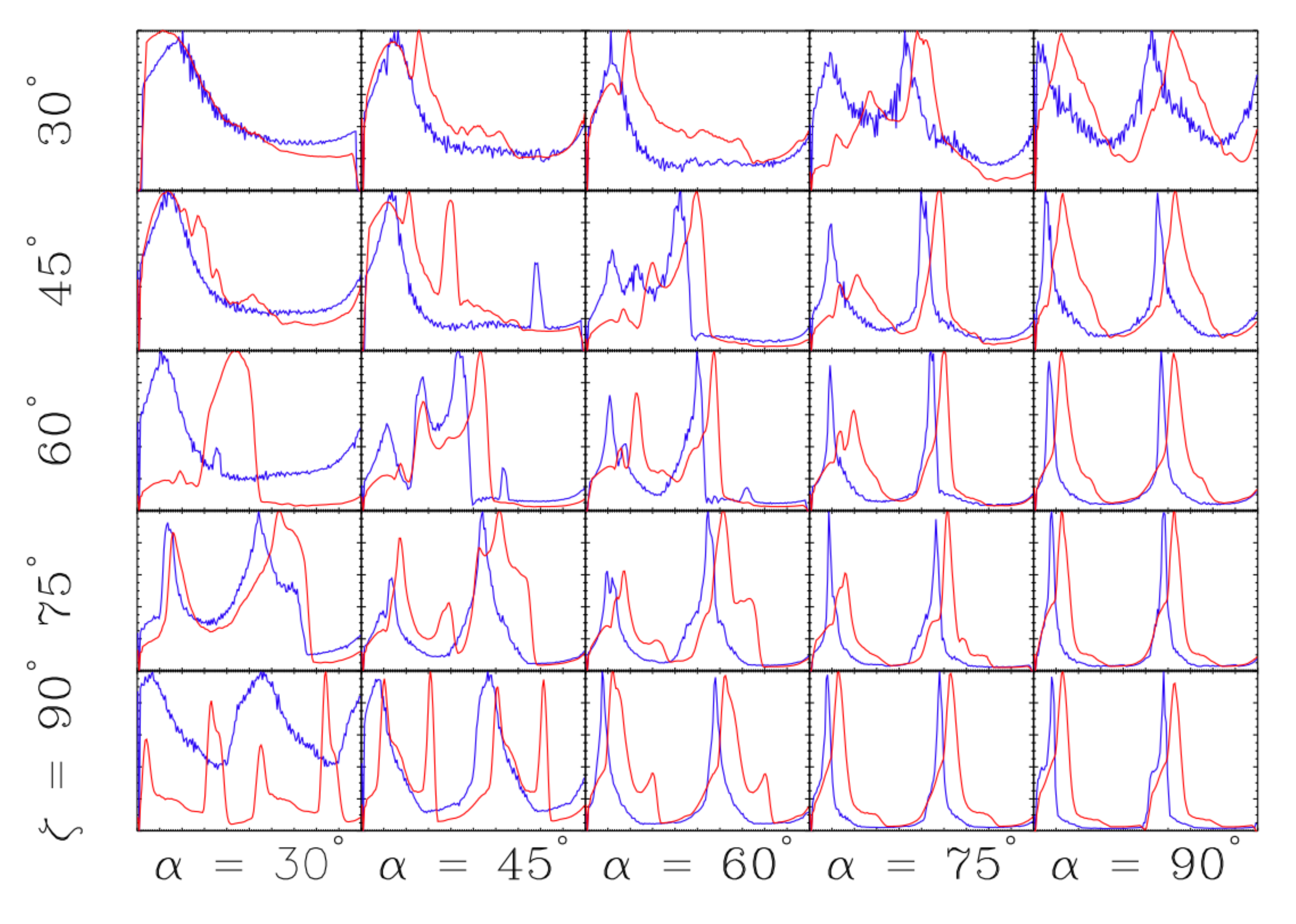}
\caption{Comparison of simulated gamma-ray light curves in retarded vacuum dipole (blue) and force-free (red) magnetospheres for asymmetric slot gap geometry.
The model parameters are the same as in Figure 2.} \label{f3}
\end{figure*}

\begin{figure*}[t]
\centering
\includegraphics[width=150mm]{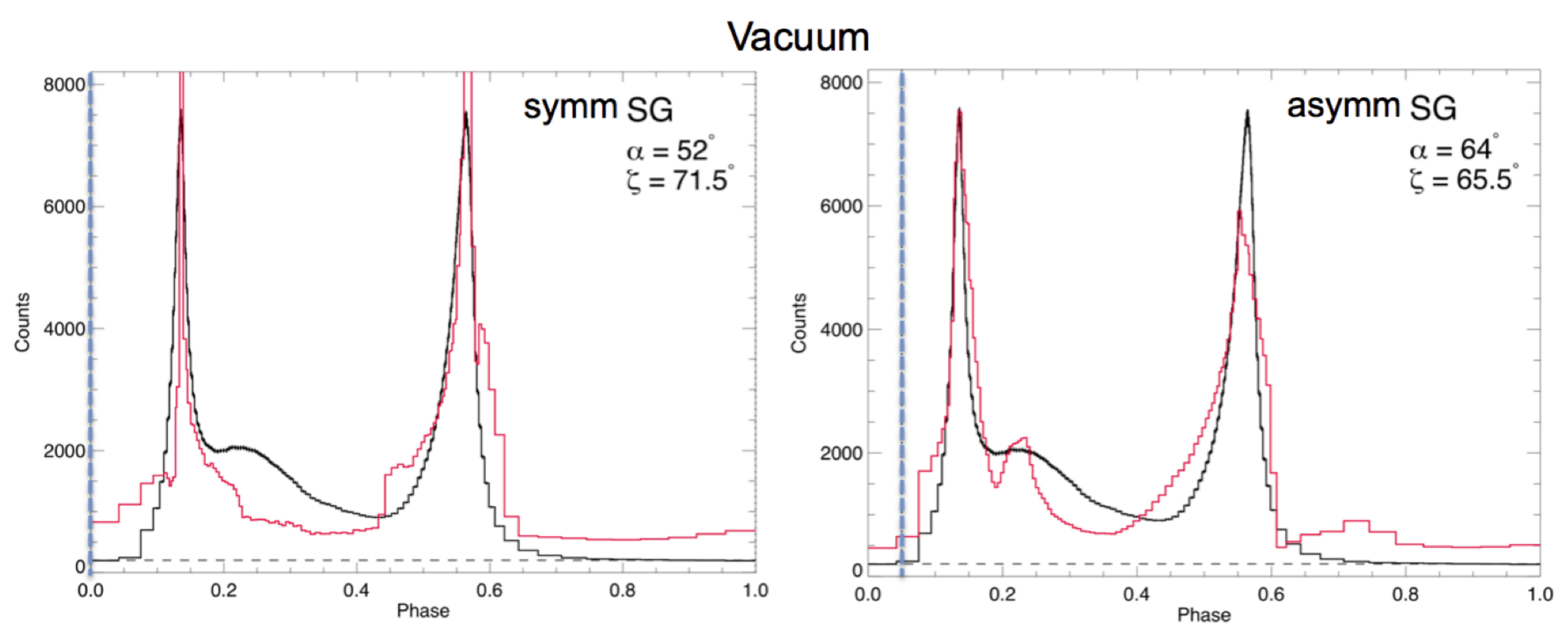}
 
\includegraphics[width=150mm]{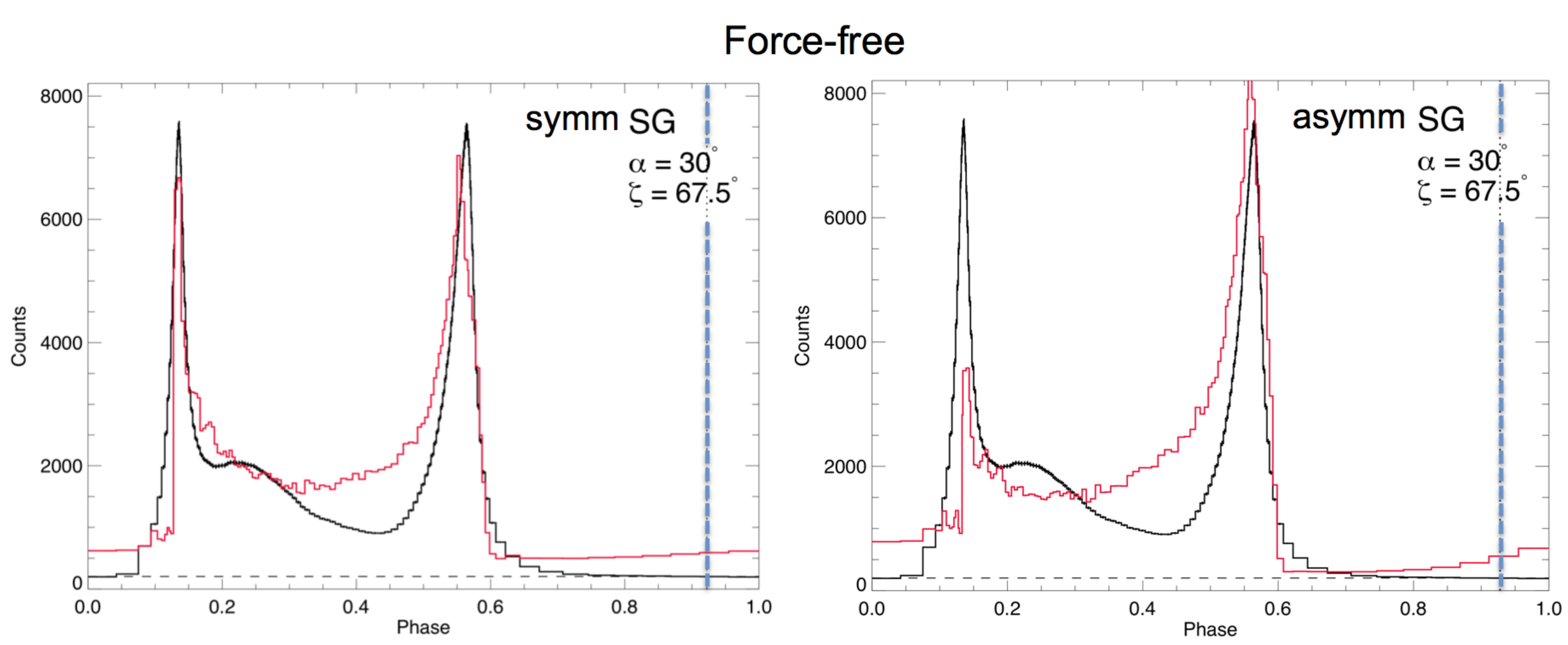}
\caption{Observed (black) and best-fit model (red) light curves for the Vela pulsar for retarded vacuum dipole (top panels) and force-free (bottom panels) 
magnetospheres.   The radio peak is at phase 0 and the vertical blue dotted lines mark the phase of the magnetic pole crossing.} \label{f4}
\end{figure*}

To simulate SG light curves in magnetospheres with offset PCs, we need to know how the asymmetry in $E_{\parallel}$ is related to a corresponding asymmetry in 
the photon emission rate.  The Lorentz factors of electrons accelerating in the SG are limited by curvature radiation reaction \cite{MuslimovHarding2004} and reach a 
steady-state
\be
\gamma_{\rm CR} = \left({3\over 2}{\rho_c^2 E_{\parallel}\over e}\right)^{1/4}
\ee
where $\rho_c$ is the local field line radius of curvature.
If the main emission from electrons accelerating in the SG comes from curvature radiation, then the photon emission rate is equal to the electron energy loss rate 
of $\dot \gamma \propto \gamma_{\rm CR}^4 \propto E_{\parallel}$.  Thus we can assume that the asymmetry in the $E_{\parallel}$ from Eqn (\ref{Epar}) produces 
a similarly asymmetric emission pattern of SG radiation.  

Figure 2 shows an atlas of SG model light curves for centered and offset PCs of the vacuum dipole,  and different values of pulsar inclination $\alpha$ and viewing 
angle $\zeta$ with respect to the rotation axis.  The light curve calculations follow that of \cite{Dyks2004}, except that here we modulate the emission in azimuth according to 
Equation (\ref{Epar}).  For 
these calculations we assume a slot gap width of $w = 0.05$ and uniform emission along field lines in the corotating frame from the neutron star surface to maximum 
radius of $r^{\rm max} = 1.2 R_{\rm LC}$, limited by a maximum cylindrical radius of $r^{\rm max}_{\rm cyl} = 0.95 R_{\rm LC}$.
The combined phase shifts from aberration and time-of-flight, of photons radiated tangent to a magnetic dipole field 
from the polar cap to the light cylinder, nearly cancel those due to field line curvature on  
the trailing edge of the open-field region.  Radiation along such trailing field lines 
bunches in phase, forming sharp emission peaks or caustics in the pulse profile.  For large $\alpha$, the caustics from both magnetic poles are visible and form two sharp 
peaks in the light curves, whose separation depends on viewing angle $\zeta$. 
On the leading side, these phase shifts add up to spread photons emitted at various
altitudes over a large range of phases, forming the bridge and off-peak emission.  From Figure 2, one can see that the light curves for centered PCs have high levels of 
off-peak emission, higher than that seen in most {\it Fermi} pulsar light curves.  However, the light curves for asymmetric SG emission resulting from offset PCs have a much lower 
level of off-peak emission, since emission is minimum on the leading edge and maximum on the trailing edge of the open field volume.  As a result, the peak level and the 
peak-to-off-peak ratio is enhanced, but does not change the phase lag of the first peak with phase 0 (left-hand plot boundary).

Figure 3 shows the $\alpha-\zeta$ atlas comparing asymmetric SG light curves for offset PC in the vacuum and force-free models for the same model parameters as Figure 2.  
The light curve morphology in the force-free 
magnetosphere is similar to that of the vacuum dipole, but the peaks are shifted in phase.  The first peak of the force-free light curves occurs at later phases relative to 
the magnetic pole crossing, 0.05 - 0.15  depending on inclination angle, compared to the vacuum light curves.
Force-free model light curves, computed by  \cite{BaiSpitkovsky2010} for the classic TPC model ($r^{\rm max}_{\rm cyl} = 0.75 R_{\rm LC}$), 
and by \cite{ContKala2010} for outer magnetosphere curvature radiation, show similar delayed phase lags of the first peaks.  The larger phase lags of the force-free model 
gamma-ray peaks result from two effects: the PCs are larger than the vacuum PCs, and the field lines are more swept back than in the vacuum dipole 
magnetosphere.  Both effects cause a larger offset toward the trailing side of the PC, delaying the caustics forming the main peaks to later phase (see Figure 1).
The radio peak, thought to come from cone beam emission centered on the magnetic pole at altitudes of several hundred km above the neutron star surface, 
would likely occur at earlier phase than that of the magnetic pole due to aberration and retardation, so the total phase lag of the first gamma-ray peak to the 
radio peak will be even larger.  

\subsection{Test Case: the Vela Pulsar}

The different shapes and radio phase lags of the gamma-ray light curves for the vacuum dipole and force-free models provide a potentially useful 
diagnostic for pulsar magnetosphere structure.  As an illustration, we have fit the Vela pulsar light curve with our vacuum dipole and force-free model 
light curves for both symmetric and asymmetric SGs (see also \cite{DeCesar2011}).  We use 30 months of sky survey data, choosing diffuse class photons within 
max$[1.6-3\log_{10}(E_{\rm GeV}), 1.3]$ degrees of the 
Vela radio position between 0.1 and 300 GeV.  The light curve is obtained by phase-folding, using the {\it Fermi} ephemeris, and then displayed in 140 phase 
bins of 3000 counts each.  We use a Markov Chain Monte Carlo maximum likelihood technique \cite{Verde2003} to determine the best fit $\alpha$, 
$\zeta$, gap width $w$, maximum emission radius $r^{\rm max}$ and magnetic pole phase lag $\Delta \phi$.  
Our resolution is $1^\circ$ in $\alpha$ and $\zeta$ for the vacuum case and $15^\circ$ in $\alpha$ (each a separate numerical solution) and 
$1^\circ$ in $\zeta$ for the force-free case.   
The resolution in gap width $0 < w < 0.3$ is 0.01 in fraction of the open volume, and resolution in maximum emission radius $0.7R_{\rm LC} < r^{\rm max} < 2.0 R_{\rm LC}$ 
is $0.1 R_{\rm LC}$. 

The fit results are shown in Figure 4, with the best fit model light curves, those with the lowest reduced $\chi_{\nu}^2 = \chi^2/135$, in red and the observed 
{\it Fermi} light curves in back.  The best-fit ($\alpha$, $\zeta$, $w$, $r^{\rm max}$, $\Delta \phi$, $\chi_{\nu}^2$)
values are ($52^\circ$, $71.5^\circ$, 0, 1.0, $0^\circ$, 1964) and ($64^\circ$, $65.5^\circ$, 0.1, 1.2, $18^\circ$, 1169) 
for symmetric and asymmetric vacuum dipole and ($30^\circ$, $67.5^\circ$, 0, 1.5, $332^\circ$, 1145) and 
($30^\circ$, $64^\circ$ 0, 1.2, $334^\circ$, 1515) for force-free symmetric and 
asymmetric models.  In the vacuum dipole model, 
the best fit ($\alpha$, $\zeta$) are different for the symmetric and asymmetric cases since their different levels of off-peak emission produce a maximum 
likelihood in an alternate location in parameter space.  In this case, the asymmetric SG model is significantly favored and has a smaller $\beta = \zeta - \alpha$
and $\Delta \phi$, allowing for some aberration of the radio peak position, which adds to the radio phase lag. The $\zeta = 65^\circ$ in this case also agrees with the viewing angle 
inferred from the X-ray torus fit \cite{NgRomani2008}.   In the force-free model, the best fit ($\alpha$, $\zeta$, $w$, $\Delta \phi$)
are the same for the symmetric and asymmetric cases, but the $r^{\rm max}$ is smaller for the asymmetric case.  In both force-free fits, the magnetic pole phase lags 
are much larger than those of the vacuum dipole models, requiring the radio peak, at phase 0 in the plot, to arrive at phase 0.08 later than the magnetic pole phase.  This would not be expected for a standard cone beam unless there was emission from only the trailing part of the cone.  Additionally, $\beta = 37.5^\circ$ in the force-free
best-fit models which is much larger than the expected width of the radio cone beam so it is not likely that the radio emission would be visible for this geometry.  

\section{CONCLUSIONS}

We have explored the gamma-ray light curves in pulsar magnetosphere geometries at the two extremes, vacuum dipole and force-free, which 
both have offset polar caps.  
In the slot gap emission model, the offset of the polar caps causes two major effects: a larger phase lag of the gamma-ray peaks with the 
magnetic pole, and a larger ratio of peak-to-off-peak emission level due to asymmetry in the slot gap emission.  The sweepback of field lines of the force-free magnetosphere is larger than
for the vacuum dipole, producing a larger polar cap offset.  The force-free model gamma-ray light curves thus show a larger radio phase lag.  
We fit simulated gamma-ray light curves using slot gap emission model 
geometry that include these effects to the observed Vela pulsar 30 month light curves.  We find that the asymmetric slot gap in the vacuum dipole case produces a 
significantly better fit to the observed light curve since the higher accelerating electric field on the trailing field lines produces a lower level of off-peak emission.  The viewing 
angle of $\zeta = 65^\circ$ for this fit also better matches the $\zeta$ from the X-ray torus fit.  We find that the force-free model light curve fit is not as favorable as that of the vacuum dipole 
in several respects.  While the symmetric SG in force-free geometry has the lowest $\chi_{\nu}^2$, the asymmetric SG fit which takes into account the offset PC is significantly worse.
The first gamma-ray peak has a phase lag of $\sim 0.2$ with the magnetic pole, which is already larger than the observed Vela radio phase lag of $0.14$, requiring emission on 
only the trailing part of a radio cone.  However, the phase of the trailing edge of a cone beam would be frequency dependent and the observed Vela phase lag is not.  By 
contrast, the gamma-ray phase lag of the vacuum dipole light curves are smaller and are consistent with viewing the radio core beam or the edge of a cone beam.
Furthermore, at the impact angle $\beta = 37.5^\circ$ of the force-free model fits, one would not expect to intercept the bright part of a radio core or cone beam.  Although
the force-free magnetosphere is a limiting case, these fits may indicate that the real pulsar magnetosphere solution is closer to the vacuum dipole in field geometry. 
Such light curve fitting that constrains the viewing geometry and radio phase lag will be an important tool in determining the geometry of the pulsar magnetosphere.

\begin{acknowledgments}
AKH acknowledges support from the {\it Fermi} GI and NASA Astrophysics Theory Programs
\end{acknowledgments}


\end{document}